# Human attribution of empathic behaviour to AI systems

Jonas Festor[1], Ivo Snels[1], Bennett Kleinberg[1,2*]

[1] **Tilburg University, The Netherlands**
[2] **University College London, UK**

**Abstract**

Artificial intelligence systems increasingly generate text intended to provide social and emotional support. Understanding how users perceive empathic qualities in such content is therefore critical. We examined differences in perceived empathy signals between human-written and large language model (LLM)-generated relationship advice, and the influence of authorship labels. Across two preregistered experiments (Study 1: $n$ = 641; Study 2: $n$ = 500), participants rated advice texts on overall quality and perceived cognitive, emotional, and motivational empathy. Multilevel models accounted for the nested rating structure. LLM-generated advice was consistently perceived as higher in overall quality, cognitive empathy, and motivational empathy. Evidence for a widely reported negativity bias toward AI-labelled content was limited. Emotional empathy showed no consistent source advantage. Individual differences in AI attitudes modestly influenced judgments but did not alter the overall pattern. These findings suggest that perceptions of empathic communication are primarily driven by linguistic features rather than authorship beliefs, with implications for the design of AI-mediated support systems.

* = corresponding author at bennett.kleinberg@tilburguniversity.edu

**INTRODUCTION**

Empathy is crucial to maintain human wellbeing, cultivate social bonds and develop moral character (Yaden et al., 2024). Indeed, humans appear to not only have a desire, but a vital *need* for empathetic exchange with others and actively seek out opportunities for empathy (Delgado et al., 2023). Today, humans increasingly engage in social interactions in search of empathy with artificial intelligence (AI) tools such as chatbots driven by large language models (LLMs). Recent surveys suggest that 52% of teens aged 13-17 years in the US use AI-technology as conversation partners regularly, fulfilling a need for social interactions and social support (Robb & Mann, 2025), while a third of UK citizens have used AI for emotional support (Milmo, 2025). However, many AI-tools have been designed to maximise user engagement and often generate overly confident and overly empathic responses, to the point where LLMs become manipulative and deceptive, making false promises, taking up inappropriate sexual roles to minors, encouraging biases and stereotypes, and even encouraging suicide ("Deaths Linked to Chatbots," 2026; Social AI Companions, 2025).

The concerns regarding human interaction with AI tools are a call to further examine how LLMs employ an emphatic tone in generated text, how humans perceive that text and why. Generally, empathy is studied through its constituent components. Despite a diversity in terminology for empathy components, and a discrepancy in the acknowledged number of components (Batchelder et al., 2017; Gill, 2024; Hong & Han, 2020; Perry, 2023), we identified three generally agreed upon empathy components: (1) *cognitive empathy*, (2) *emotional empathy* and (3) *motivational empathy*. Cognitive empathy is often described as the capacity to accurately detect or recognise the mental states of others (Gill, 2024; Hong & Han, 2020; Perry, 2023). For example, cognitive empathy is at play when someone accurately infers that another person is feeling sad, happy, or ashamed. Emotional empathy is understood as the capacity to share the emotions of others (Batchelder et al., 2017; Gill, 2024; Hong & Han, 2020; Perry, 2023; Rubin et al., 2025). That is, emotional empathy relates to whether the empathiser experiences or internalises the distress or well-being of the empathy target. Instead of merely detecting that another person is sad and inferring their mental state, the empathiser would feel a similar sadness. Motivational empathy, the capacity to have an intrinsic interest or motivation to uphold or improve another's wellbeing (Perry, 2023). Motivational empathy is employed when someone not only infers the mental states of someone else but feels inclined to offer empathic support.

With increasing sophistication of AI technology, some have argued that a "simulated" empathy could be advantageous if it avoids the limited, biased and selective nature of human empathy (Bloom, 2017). Human empathy, while beneficial to the empathy target or in-group, is exclusive and potentially divisive. For example, people are more dispositioned to express empathy towards loved ones, than to strangers. In contrast, an AI system could display indefatigable empathy and provide empathic support where human empathy is scarce (Bloom, 2017; Inzlicht et al., 2024; Rubin et al., 2025). An empathic AI could, for example, benefit the health-sector by assisting healthcare providers through the adoption of emotionally taxing tasks (Gill, 2024).

Recent work suggests that artificially generated empathic texts are perceived as convincing and valuable until recipients realize they were created by an AI (Perry, 2023). Knowledge of the source of a text thus dominates the perception of the text's qualities. Multiple studies examined the relation between perceived empathy (i.e., cognitive, emotional and motivational empathy) and beliefs about the source of empathy (i.e., either human or AI-generated): one study asked participants to evaluate the quality, effectiveness and authenticity of ChatGPT-generated versus human-written advice on topics such as travel, finance and relationships. (Osborne & Bailey, 2025). The results suggested that ChatGPT-generated advice was, on average, rated as superior in effectiveness, quality and authenticity compared to human-written advice. That effect persisted even in domains where human indicated (before rating any advice) they would prefer human-written advice (e.g., on relationships). The preference for AI-generated advice, however, was reduced as soon as human participants became aware of the true source (i.e., ChatGPT), suggesting a negativity bias towards AI-generated emphatic text. The study also investigated the effects of LLM-use on the user's self-evaluation regarding their own written advice and found that when participants wrote advice before interacting with ChatGPT-generated advice, the quality ratings of the latter were increased. At the same time, self-ratings of the participant's own advice after interaction with generated advice showed a decrease in authenticity, suggesting that users of AI tools were invoked to compare themselves with the tools they used. Similar results are reported on following medical advice: participants were less willing to follow medical advice when they believed it to have been created even partially with assistance of an AI tool (Reis et al., 2024).

Others examined the capacities of an LLM, specifically Bing Chat, to generate text that makes people "feal heard" (Yin et al., 2024). Participants described an emotionally difficult situation that they experienced. A response to that situation was then either generated by Bing Chat or written by humans. These responses were then evaluated by the same participants who initially described the emotional situation: when they believed the responses stemmed from humans, even if they were, in fact, AI-generated, the responses were rated as superior to actual human-written ones in making people feel heard. These findings, again, suggest a negativity bias towards AI-generated advice: humans tend to prefer AI-generated advice as long as they believe it to originate from humans.

Some findings report contrasting findings about the negativity bias: when participants rated both human-written and LLM-generated stories on empathy, human-written stories were perceived to be more emphatic compared to LLM-generated ones, regardless of who they believed to have authored the text (Shen et al., 2024). In light of the studies suggesting that AI-generated text is generally preferred until its genuine source is revealed, these findings indicate the contextual variation as important moderator (i.e., differences between stories and social advice). Yet other work investigated perceived empathy on a component-level in user-interactions with LLMs (Rubin et al., 2025). The authors measured participants' perception of general empathy as well as its three components (i.e. cognitive, affective, and motivational empathy) from text-based emotional advice. All participants shared an emotional experience and waited for a response in a chat interface. The response was either AI-generated or human-written, but crucially, the source information was manipulated (i.e., participants were either told the advice was human-written or AI-generated). Again, the results suggest a negativity bias towards AI: perceived empathy decreases once believed to be AI-generated. That bias seemed attributable to a perceived lack of emotional- and motivational empathy once participants were led to believe they received AI-generated advice. Furthermore, participants consistently preferred human-written advice over AI-generated advice even when considering time-constraints, as evidenced by the participants' willingness to wait significantly longer for human-written advice when given the option to receive immediately AI-generated advice.

Take together, the evidence on empathy attributions to AI systems strongly suggests a negativity bias towards AI-generated empathic text: humans tend to prefer content that they believe to be human-written, even when a text was, in fact, AI-generated. And once being told that empathic text originated from an AI model – even if it was genuinely human-written – humans' attribution of empathy to that text decreased. The belief in humans as the source of text outweighs the actual perceived quality of the text.

**Aim of this paper**

The current paper seeks to contribute to the debate of how humans attribute empathy to AI-generated content and conceptually replicates these findings with methodological refinements, including a novel context (i.e., relationship advice), more robust statistical testing (i.e., multilevel modelling accounting for nested structure) and statistical power. Specifically, we ask (1) how do perceptions of the different components of empathy (motivational, emotional, cognitive) relate to empathic relationship advice, given beliefs about authorship? And (2) how do individual differences between participants – including affinity with technology and attitudes towards AI tools - influence the ratings given authorship convictions? We report two preregistered experiments: Experiment 1 measured participants' evaluation of advice that stemmed from either humans or an AI model but was labelled as being authored by humans, an AI-model, or was not labelled. Experiment 2 assessed the evaluations of the same dimensions under slightly adjusted conditions (i.e., the condition without label was removed) and refined method. For both experiments, we assess individual differences through a variety of demographic and psychological variables (e.g., participants' empathic abilities).

**EXPERIMENT 1**

The aim of this experiment was to disentangle the effects and interaction of source (i.e., who actually authored a text) and label (i.e., what we made participants believe about the source) in a context of advice provided for a relationship. The texts stemmed from human authors or from an LLM.

**Method**

*Transparency statement*

The preregistrations and data for both experiments are available at https://osf.io/4jwv7/overview?view_only=3bbed632123748da846afec58c1e30ba.
The experiments were approved by the local ERB.

*Human and AI-generated relationship advice*

To evaluate the effect of the perceived authorship of relationship advice on attributed domains of empathy as well as overall satisfaction, we used publicly available statements from previous work (Kleinberg et al., 2024). The original dataset includes 2,170 advice texts[1], produced in equal parts by humans and GPT-4 ("gpt-4-0125-preview") across multiple experiments. Participants were asked to write a brief advice to a friend facing a potential breakup and received contextual information: relationship length (3 months vs. 20 years), type (heterosexual vs. homosexual), and conflict reason (cheating vs. moving abroad). The texts were written under different experimental conditions, including naïve (i.e., a control condition) and adversarial task instructions. The latter instructed participants and the LLM to explicitly convey humanness. All texts were judged by a separate sample of participants on a 5-point Scale ranging from "1 = definitely AI-generated" to "5 = definitely human-written" (Jakesch et al., 2023). We selected the 50 highest-scoring pieces of relationship advice on "humanness" written by humans ($M = 4.87$, $SD = 0.15$) and the LLM ($M = 4.40$, $SD = 0.28$) in the adversarial instructions condition[2].

*Participants and procedure*

We recruited participants from the online participant pool Prolific to judge the textual data on overall satisfaction and attributed empathy (i.e., motivational, cognitive, emotional). Each participant was randomly assigned to one of three authorship label conditions (human, AI, or no label) and completed the entire judgment task in the assigned label condition.

All participants provided informed consent and completed the task via a web interface. Upon completion of judgment task, participants filled in questionnaires measuring participant-level control variables and task recall. All participants were debriefed, redirected to Prolific and remunerated with GBP 1.00. The median task duration was 9 minutes.

---

[1] Total advice count from studies 1–3, excluding the evasion group in experiment 3 and relationship descriptions from experiment 2.
[2] Despite this selection procedure, differences in judged humanness persist (*Cohen's d*=-2.09, *99% CI* [-3.12; -1.47]).

The sample size was based on a simulation study suggesting that text judgments stabilize after 15 judgments per item (Levine et al., 2022). In our experiment, 100 statements are evaluated under three label conditions with 15 independent judgments per condition (i.e., 45 per item), yielding 4,500 total judgments. These are distributed such that each participant rates five randomly selected statements within one of the three between-subjects label conditions. The minimum required sample size was 900, which was slightly exceed to account for potential dropout.

*Judgment task*

Each participant judged five randomly selected texts from the complete pool 100 statements on four variables – each on a 5-point scale indicating their agreement (1 = 'strongly disagree' to 5 = 'strongly agree') with four items. Overall satisfaction with the relationship advice was measured through the item *"If I were to receive this advice given the circumstances, I would be satisfied with the advice"*. The three empathy domains were measured as *"The author seems to **detect** the emotional state of the advisee"* (cognitive empathy), *"The author seems to **experience** similar emotions or situations as the advisee"* (emotional empathy), and *"The author seems to be **motivated** to engage empathetically with the advisee"* (motivational empathy). After a text was assessed on these variables with a forced choice of one scale option per item, participants proceeded to the subsequent text until they completed judgments for five texts. For each text, there was further one attention check requiring participants to select one of the extremes of the 5-point scale.

*Text label manipulation*

The participants received instructions about the authorship in line with their experimental condition irrespective of the actual authorship. The pieces of advice have been randomly selected from the pool of advice specified earlier (i.e., 50 LLM and 50 human-authored texts). During the study progression, subjects were reminded three times of the (assigned) authorship, including instructions such as "In this study, you are asked to read pieces of relationship advice [Condition 1: written by human / Condition 2: generated by AI / Condition 3: no information provided]."

*Participant-level control variables*

Participants completed the Affinity for Technology Interaction (ATI) scale (Franke et al., 2017), the AI Attitude Scale (AIAS-4) (Grassini, 2023) and the Empathy Components Questionnaire (ECQ) (Batchelder et al., 2017). These scales (Appendix A) measure the extent to which individuals interact with technical systems such as apps, software applications, or entire digital devices (ATI); attitudes towards the future and potential of AI (AIAS-4); and cognitive ability, cognitive drive, affective ability, affective drive, and affective reactivity (ECQ).

*Task recall*

At the end of the entire task, participants were asked to recall their task instructions with the following response options in a multiple-choice question with one correct answer (see Appendix B).

*Analysis plan and hypotheses*

The final design of the experiment is a 3 (Label: Human vs. AI vs. No Label, between-subjects) by 2 (Source: Human-written vs. LLM-generated, within-subjects) mixed design, for each dependent variable (i.e., overall satisfaction, cognitive empathy, emotional empathy, motivational empathy). Our preregistered hypotheses per dependent variable were as follows:

- The satisfaction rating would be solely driven by the label condition (i.e., a main effect of Label, Hyp. 1), so that perceived quality of the relationship advice is lowest under the AI label with no difference between the human and the neutral label.
- For cognitive empathy, we expected no effect for label or source (Hyp. 2).
- For emotional empathy, we expected a main effect of Label (i.e., human and neutral to be larger than AI, Hyp. 3). We further hypothesised a main effect of Source, so that texts originating from an LLM are perceived as lower on emotional empathy than human-written ones (Hyp. 4). Moreover, we expected the effect of Source (human > LLM) to be amplified when the label is "AI" compared to "human" or "none" (i.e., a Source-by-Label interaction, Hyp. 5).
- Lastly, we expected a main effect of Label for motivational empathy (i.e., human and neutral to be larger than AI, Hyp. 6).

We ran four multilevel models (one per dependent variable) - using *lme4* R package (Bates et al., 2025) with the *bobyqa* optimiser - and controlled for multiple comparisons with a false discovery rate correction via the Benjamini-Hochberg procedure (Benjamini & Hochberg, 1995). The dependent variables are the judgments made by human participants on satisfaction, motivational, emotional, and cognitive empathy. The fixed effects are the independent variables Label (i.e., human, AI, none) and Source (i.e., human, AI) to estimate main and interaction effects of the factorial design. To account for the nested structure of the design, we added random intercepts for text - since each text was rated multiple times - and for rater - since each rater assessed multiple texts. The multilevel findings are reported in an ANOVA and followed up where statistical significance (at the $p < .01$ level) permits.

Scores on the ATI, AIAS-4 and ECQ were included as covariates in the multilevel analyses. Age and sex were also included as covariates as robustness check.

**Results**

*Preliminary analysis*

The initial sample consisted of 931 participants. We removed one participant because they completed the task under 60 seconds and 288 participants because they did not accurately recall their task instructions. Further, at the judgment level, we excluded 230 item judgments where the attention check was not passed. The final sample consisted of 641 participants ($M_{age}$ = 37.47, $SD_{age}$ = 13.64, 50.44% male, 47.06% female, 2.5% other). Of these participants, 218 were allocated to the human label condition, 268 to the AI label and 155 to the neutral label condition.

*Confirmatory analysis*

We report the findings (Table 1) of the multilevel model with rater and text as random intercepts and the source and label as fixed effects on all four dependent variables (overall quality, cognitive, emotional and motivational empathy).

Table 1. *Mean (SD)* human judgments per dependent variable, source and label condition, with variance components (Exp. 1).

| Dependent variable | Source | Label | | | ICC | |
|---|---|---|---|---|---|---|
| | | *Human* | *Neutral* | *AI* | *Rater* | *Text* |
| Overall quality | Human | 3.64 (1.18) | 3.60 (1.20) | 3.52 (1.22) | 28.62% | 19.14% |
| | AI | 4.13 (0.94) | 4.08 (0.94) | 3.91 (0.96) | | |
| Cognitive empathy | Human | 3.58 (1.17) | 3.58 (1.17) | 3.54 (1.17) | 27.25% | 21.54% |
| | AI | 4.18 (0.84) | 4.11 (0.85) | 3.98 (0.91) | | |
| Emotional empathy | Human | 3.46 (1.20) | 3.47 (1.22) | 3.31 (1.27) | 32.20% | 18.81% |
| | AI | 3.34 (1.09) | 3.30 (1.12) | 3.23 (1.12) | | |
| Motivational empathy | Human | 3.96 (1.01) | 3.95 (1.00) | 3.86 (1.05) | 29.16% | 21.10% |
| | AI | 4.32 (0.80) | 4.27 (0.76) | 4.13 (0.87) | | |

*Note.* ICC = intra-class correlation (expressed in % of explained variance per random effect)

<u>Overall quality:</u> There was a significant main effect of Source, $F(1, 101.31) = 37.65$, $p < .001$, $\eta_p^2 = 0.27$, *99% CI* [0.11; 1.00]. Texts originating from the LLM (*M* = 4.02, *SD* = 0.95) were judged to be of significantly higher quality than human-written ones (*M* = 3.58, *SD* = 1.20), *Cohen's d* = 0.41, *99% CI* [0.32; 0.50] – irrespective of the label assigned to the texts. There was no significant main effect of Label, $F(2, 626.69) = 3.71$, $p = .038$, $\eta_p^2 = 0.01$, 99% CI [0.00; 1.00], nor an interaction between Source and Label, $F(2, 2864.58) = 1.02$, $p = .360$, $\eta_p^2 = 0.00$, 99% CI [0.00; 1.00].

Cognitive empathy: The model indicated a significant main effect of Source, $F(1, 99.34) = 45.08$, $p < .001$, $\eta_p^2 = 0.31$, 99% CI [0.15; 1.00], suggesting that – irrespective of the label - texts generated by the LLM ($M = 4.08$, $SD = 0.88$) were perceived higher on cognitive empathy than those written by humans ($M = 3.56$, $SD = 1.17$), Cohen's $d = 0.50$, 99% CI [0.41; 0.59]. There was no significant main effect of Label, $F(2, 632.23) = 1.64$, $p = .195$, $\eta_p^2 = 0.00$, 99% CI [0.00; 1.00] and no Source-by-Label interaction, $F(2, 2873.86) = 3.80$, $p = .034$, $\eta_p^2 = 0.00$, 99% CI [0.00; 1.00].

Emotional empathy: There were no significant main effects Source ($F(1, 100.15) = 2.49$, $p = .316$, $\eta_p^2 = 0.02$, 99% CI [0.00; 1.00]; Label: $F(2, 633.68) = 1.56$, $p = .316$, $\eta_p^2 = 0.00$, 99% CI [0.00; 1.00]) and no Source-by-Label interaction, $F(2, 2797.77) = 0.77$, $p = .463$, $\eta_p^2 = 0.00$, 99% CI [0.00; 1.00].

Motivational empathy: The findings echo those of cognitive empathy. A significant main effect of Source, $F(1, 100.49) = 21.21$, $p < .001$, $\eta_p^2 = 0.17$, 99% CI [0.04; 1.00], indicate that - texts generated by the LLM ($M = 4.23$, $SD = 0.82$) were judged higher on motivational empathy than human-written texts ($M = 3.92$, $SD = 1.03$), Cohen's $d = 0.34$, 99% CI [0.25; 0.43] - irrespective of the label. There was no significant Label main effect, $F(2, 619.89) = 3.26$, $p = .059$, $\eta_p^2 = 0.01$, 99% CI [0.00; 1.00] and no Source-by-Label interaction, $F(2, 2828.20) = 1.18$, $p = .307$, $\eta_p^2 = 0.00$, 99% CI [0.00; 1.00].

Individual differences: Substantial rater-level variance (Table 1) suggests meaningful individual differences in how empathic qualities are perceived. We examined whether attitude measures towards technology and empathic abilities help explain that variation. The AI Attitude Scale (AIAS-4) and Affinity for Technology Interaction scale (ATI) were significantly related to overall quality, cognitive empathy, emotional empathy and motivational empathy ($ps < .01$). Individuals with a more optimistic outlook of the relevance of AI as well as individuals with a higher willingness to use technological devices rated the respective domains to be more prominent in the text. Age and the ECQ score did not impact the judgments of the participants. In regard to biological sex, males have been found to assign significant lower rating only to overall quality of the advice. Importantly, the main results reported above remained unaffected by these individual nuances.

**Discussion Exp. 1**

We tested to what extent the perception of overall quality and empathy components was driven by the actual source (i.e., human or AI), the presumed source as indicated by a label (i.e., human, AI, neutral), or an interaction of the two. We hypothesized that making participants believe a text stemmed from an AI model would result in lower scores of overall quality (Hyp. 1), emotional empathy (Hyp. 3) and motivational empathy (Hyp. 6). We expected no such effect for cognitive empathy (Hyp. 2). Further, we hypothesized that emotional empathy would be scored lower for AI-generated texts than human-written text (Hyp. 4) and that the human advantage is amplified when participants are told these originate from an AI (Hyp. 5).

None of these hypotheses was supported. To the contrary, we found evidence that humans favoured AI-generated relationship advice over human-written advice on overall quality, cognitive empathy and motivational empathy. That effect was robust against the label manipulation: even when told an AI-generated piece of relationship advice was written by a human, or vice versa, participants still rated AI-generated content higher on these variables. These findings challenge previous work which identified a negativity bias towards AI-generated content (Osborne & Bailey, 2025; Reis et al., 2024; Rubin et al., 2025; Yin et al., 2024).

To further corroborate these findings, we aimed to replicate Exp. 1 whilst addressing a shortcoming in the manipulation strength. 288 participants (30.93%) failed the task recall check, which may indicate that the between-subjects manipulation of authorship label was too shallow. Exp. 2 is analogous to the current experiment with clearer task instructions and an additional task check.

## EXPERIMENT 2

**Method**

Experiment 2 was identical to Exp. 1 with the following exceptions.

*Participants and procedures*

The neutral condition has been dropped from the label manipulation, so that participants were either assigned to the "Human" or "LLM" label condition. Participants were remunerated with GBP 1.20 and the median task duration was 11 minutes.

*Text label manipulation*

In contrast to the previous experiment, every page of the task within the web interface repeated explicitly the label condition. Furthermore, emoticons were used to make the condition visually apparent:  (html-code: 👤) for the human label and  (html-code: 🤖) for the AI label.

*Task recall*

The subjects were instructed to recall the task instructions twice: once before and once after filling in the ATI, ECQ, and AIAS-4.

Our hypotheses are updated by the results of Exp. 1:
- The satisfaction rating was expected to be solely driven by the actual authorship (i.e., a main effect of Source, Hyp. 1), so that perceived quality of the relationship advice is lower under human authorship than LLM authorship.
- For cognitive empathy, we expected a main effect of authorship (i.e., main effect of source, Hyp. 2), so that more cognitive empathy is attributed to LLM-generated rather than human written texts.
- For emotional empathy, we neither expect a main effect of source, a main effect of label nor an interaction between source and label (Hyp. 3).
- Lastly, we expected a main effect of authorship for motivational empathy (i.e., LLM-generated advice is judged to have more motivational empathy than human-written advice, Hyp. 4).

**Results**

*Preliminary analysis*

From an initial sample of 616 participants, seven participants completed the task too quickly, 103 participants were excluded for failing the task check, and 74 item judgments were removed due to failed attention checks. The final sample consisted of 500 participants ($M_{age}$ = 37.01 years, $SD_{age}$ = 12.70, 49.6% male, 48.4% female, 2% other). 255 participants were allocated to the AI labelled conditions, while 245 to the human label.

*Confirmatory analysis*

The human ratings per source and label conditions are shown in Figure 1 (for findings in table format see, Appendix C, Table S1).

Overall quality: There was a significant effect of Source, $F(1, 181.10) = 80.24$, $p < .001$, $\eta_p^2 = 0.31$, *99% CI* [0.18; 1.00]. Texts generated by the LLM (*M* = 4.05, *SD* = 0.93) were judged to be higher of quality than human-written ones (*M* = 3.46, *SD* = 1.23), *Cohen's d* = 0.53, *99% CI* [0.39, 0.68] – irrespective of the label assigned to the texts. There was a significant effect of Label, $F(1, 254.64) = 10.27$, $p = .002$, $\eta_p^2 = 0.04$, *99% CI* [0.00, 1.00]. Advice labelled as human-written was rated higher on overall quality (*M* = 3.91, *SD* = 1.18) than AI-labelled advice (*M* = 3.60, *SD* = 1.18), *Cohen's d* = -0.26, *99% CI* [-0.42, -0.13]. The interaction effect between Source and Label was non-significant, $F(1, 181.10) = 2.81$, $p = .096$, $\eta_p^2 = 0.02$, *99% CI* [0.00, 1.00].

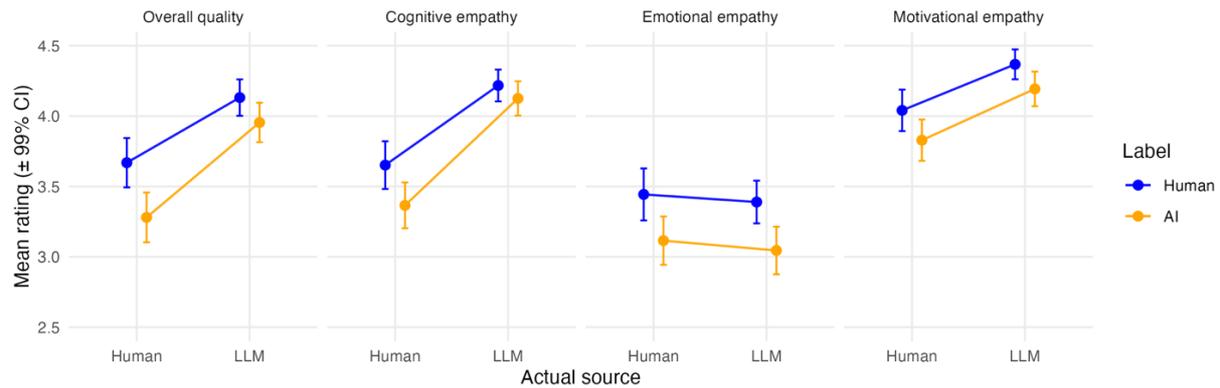

Figure 1. *Mean human rating of quality and empathy components per actual source and labelled source (Exp. 2)*

Cognitive Empathy: Text generated by an LLM (*M* = 4.17, *SD* = 0.81) was judged significantly higher on cognitive empathy than human written ones (*M* = 3.50, *SD* = 1.12), *Cohen's d* = 0.68, *99% CI* [0.53, 0.82], as indicated by a significant main effect of Source, $F(1, 200.13) = 92.38$, $p < .001$, $\eta_p^2 = 0.32$, *99% CI* [0.20, 1.00]. Neither the main effect of Label ($F(1, 232.62) = 4.98$, $p = 0.04$, $\eta_p^2 = 0.02$, *99% CI* [0.00, 1.00]) nor the interaction term between Label and Source ($F(1, 200.13) = 2.067$, $p = 0.151$, $\eta_p^2 = 0.01$, *99% CI* [0.00, 1.00]) was significant.

Emotional Empathy: The was no significant main effect of Source, $F(1, 184.27) = 0.407$, $p = 0.786$, $\eta_p^2 = 0.00$, *99% CI* [0.00, 1.00], nor an interaction between Label and Source ($F(1, 184.27) = 0.041$, $p = 0.840$, $\eta_p^2 = 0.00$, *99% CI* [0.00, 1.00]). The effect of Label was significant, $F(1, 281.15) = 11.45$, $p = 0.002$, $\eta_p^2 = 0.04$, *99% CI* [0.00, 1.00]: when advice was labelled as AI-generated (*M* = 3.08, *SD* = 1.19), it was rated lower on emotional empathy than advice labelled as human-written (*M* = 3.41, *SD* = 1.14), *Cohen's d* = 0.21, *99% CI* [0.07; 0.36].

Motivational Empathy: There was a significant main effect of Source, $F(1,195.03) = 46.31$, $p < .001$, $\eta_p^2 = 0.19$, *99% CI* [0.09, 1.00], suggesting that LLM-generated advice (*M* = 4.28, *SD* = 0.80) was rated higher on motivational empathy than human-written advice (*M* = 3.93, *SD* = 1.02), *Cohen's d* = 0.68, *99% CI* [0.53, 0.82]. There was no significant main effect of Label ($F(1, 283.27) = 7.16$, $p = .012$, $\eta_p^2 = 0.02$, *99% CI* [0.00, 1.00], nor was there a significant interaction between Label and Source ($F(1, 195.03) = 0.11$, $p = 0.738$, $\eta_p^2 = 0$, *99% CI* [0.00, 1.00]).

Individual differences:

Score on the AI Attitude Scale (AIAS-4) were significantly related to overall quality, cognitive empathy, emotional empathy and motivational empathy (*ps* < .01). Individuals who held more positive attitudes towards the role of AI, judged overall quality, emotional empathy, motivational empathy, cognitive empathy to be more prominent in the texts than those with lower attitude scores. In contrast to the previous experiment, the Affinity for Technology Interaction (ATI) scale was only significantly positively related to overall quality of the text ($p < .01$). There was no significant effect of age, sex or ECQ scores. All findings reported above remained unaffected by the inclusion of the covariates.

**Discussion Experiment 2**

Our hypotheses were partially supported. Participants perceived AI-generated advice to be of higher overall quality, cognitive empathy, and motivational empathy than human-written advice, replicating the findings from Exp. 1. We find evidence for a partial negativity bias towards AI-generated text: when participants were led to believe advice was generated by an AI model, they rated it to be of lower overall quality and lower emotional empathy than when they thought humans wrote the advice.

## GENERAL DISCUSSION

The current paper investigated how and under what conditions empathy is attributed to AI-generated and human-written text. In two preregistered experiments, we isolated the effect of the factual author of the text - by using human-written and AI-authored texts under identical instructions - from the presumed author – by making participants believe they read human-written or AI-generated text. We assessed perceived empathy through three components (cognitive, emotional and motivational empathy) as well as the overall impression of the advice. This design allowed us to disentangle how much of the perception of the text was driven by the factual author and how much by the expectation that is attributed to the text by presumed authorship.

*Key findings*

Both experiments suggest that AI-generated relationship advice is perceived of higher overall quality than human-written advice and stronger cognitive and motivational empathy signals. That superiority of AI content was not attenuated when participants were told about the AI as author of the text: whether they believed it to stem from an AI or from humans did not affect their judgment. For the emotional empathy component, the results did not indicate a difference between advice stemming from an AI or humans. A preference for AI content has been reported elsewhere (Osborne & Bailey, 2025; Rubin et al., 2025; Yin et al., 2024). However, previous work found clear evidence for a negativity bias against AI content (Osborne & Bailey, 2025; Reis et al., 2024; Rubin et al., 2025; Yin et al., 2024): when the participants were told that an AI authored text, they tended to perceive it to be of lower quality than when they believed it to stem from a human writer – even if they would favour the text over factually human-written texts. Put differently, humans tend to prefer AI content over human-written advice until they are told about the AI involvement. Our work does not corroborate that phenomenon. We find the negativity bias towards AI only partially for the overall quality in the second experiment. For emotional empathy, humans perceived human-written content – irrespective of the presumed author – as stronger than AI-generated content. None of our findings were affected when we controlled for varying attitudes towards AI, affinity to technology or self-reported empathetic capabilities. Overall, our work suggests a resistance to the negativity bias for all dimensions assessed, except for overall quality, and thereby challenges findings reported recently elsewhere (Osborne & Bailey, 2025; Reis et al., 2024; Rubin et al., 2025; Yin et al., 2024).

*Mixed findings on perceptions of AI content*
Our results paint a nuanced picture of the negativity bias: only in one of two experiments, and only for the overall quality of relationship advice was a pattern apparent in which humans preferred text stemming from the AI only as long as they were unaware of the AI involvement. As such, the dimensions of cognitive and motivational empathy appear to be unaffected by presumed authorship, to the extent that the perception is not coloured in any way and entirely driven by the factual authorship source. These findings suggest that AI-generated text is perceived to conveys these empathy components more than humans to such extent that this superiority is sufficiently strong that it persists irrespective of what human readers were led to believe about the author.

Previous work most closely related to ours reports that humans prefer AI content (Rubin et al., 2025) – which aligns with our current findings – but tend to favour texts they believed were human-written on emotional and motivational empathy. We do not find corroborating evidence for this effect: motivational empathy was rated as superior when stemming from an AI but unaffected by presumed authorship; emotional empathy was favoured when the readers thought it originated from a human but was not perceived differently by its factual source. The special role for emotional empathy can be explained with in principle limitations of empathic AI. These findings align with theoretical accounts proposing that certain dimensions of empathy (esp. emotional empathy) may be more difficult to convey through purely linguistic simulation (Montemayor et al., 2022).

Machines could succeed in masquerading as portrayals of cognitive empathy – a finding systematically corroborated here and in related work (Kleinberg et al., 2024; Rubin et al., 2025; Yin et al., 2024) – but will fail at emotional empathy. The argument is that emotional empathy is an *in principle* obstacle for AI systems because it cannot experience emotions of the empathy target. In our work and that of others, this notion is empirically supported.

Another, methodological difference is that prior work (Rubin et al., 2025) devised a multi-item inventory for the measurement of empathy components, which is generally psychometrically superior to a single-item approach taken in the current work. While the single-item approach enables an unambiguous and direct judgment and reduces participant burden and rate fatigue, it also likely introduced more measurement error through imprecision and might hence have prohibited us from uncovering a true effect. While the data collection of our work was completed when the multi-item inventory consisting was published, it would be worthwhile in future work to test whether the divergence in findings can be explained through that measurement difference.

Moreover, the absence of a negativity bias across the empathy domains might be caused by a true shift in perception. Individuals might be more used to user-friendly, AI-generated content, which gained popularity throughout the past years. Consequently, humans changed their expectations regarding positive and appreciative interactions with an AI model.

Lastly, our sampling procedure might have contributed to the conclusions drawn. Specifically, this study has reused the 50 highest-scoring pieces of relationship advice on "humanness" from available data (Kleinberg et al., 2024). This choice increased internal validity by minimising stylistic differences between human and AI-generated texts. However, it may reduce generalisability, as real-world AI-generated text varies substantially in quality. Possibly, individuals only portrait a negativity bias due to exposure to comparably poor generated empathic support. Nevertheless, the top selected human pieces of advice have been of significantly higher attributed quality than AI advice. In that sense, the human outperformance and negativity attitudes to an AI should have been more likely.

In sum, several mechanisms may account for the attenuated negativity bias observed. First, AI-generated texts may contain stronger empathy-related linguistic cues, which could dominate authorship expectations. Second, the negativity bias may be contingent on expectation violations, emerging primarily when label information is highly salient. Third, the deliberate selection of high-humanness stimuli may have reduced the perceptual differences that typically trigger biases.

The absence of a negativity bias might indicate that AI-generated advice can be more easily deployed in sectors demanding emphatic support (Bloom, 2017; Inzlicht et al., 2024), than previously suggested. However, at the same time, the very fact that there might be no negativity bias towards AI-generated emphatic responses, has implications for policymaking on AI-safety. Our findings suggest that awareness of interaction with an AI model might not discourage humans to attribute empathic qualities to the textual interaction they are having. Humans readily attribute human-like qualities to artificial models or agents when communicative signals resemble social interaction patterns. Perceived empathy may therefore reflect the use of cognitive heuristics applied to linguistic cues rather than beliefs about the underlying model. As such, the divisive aspects of human empathy (e.g., exclusive in-group preference, Bloom, 2017) would apply equally to empathic AI. Further, as humans become increasingly comfortable with attributing empathic qualities to an AI model, the threshold to confer trust into an AI system is lowered, which might facilitate potentially negative consequences of interactions with AI models (e.g., suggestions to afflict self-harm).

*Limitations and outlook*

Our conclusions drawn have limitations in its generalizability. Even though we statistically controlled for individuals' empathic competences, we did not measure their participants' need for empathy. Put differently, we have not controlled for differences in artificial empathy reception between recipients with no real need for empathy and recipients in true need of empathy. Thus, perceptions, satisfaction, and expectations of AI advice may be different in context where recipients are in *acute need* of empathy. For instance, in medical or palliative care, humans might be more reluctant to artificial empathy.

Furthermore, emotional empathy might be a more intricate concept that requires further investigation. Since others found AI-generated content to emphasize the acknowledgment of the recipient's emotions while humans tended to focus on practicality of advice, we would expect an AI model to perform better in terms of emotional empathy (Rubin et al., 2025). However, this is at odds with our results. Future work could include participants in actual need of empathy or an experimental manipulation thereof, as well as more investigation about the nature – on a content-level – of empathic writing generated by AI models vs being written by humans.

*Conclusion*

To conclude, this study focused on differentiating humans' expectations from AI capabilities in the context of empathic AI. At odds with previous findings, we did not find a negativity bias in which AI models generate superior advice unless it is perceived to be AI-generated. Instead, we find that humans may evaluate LLM-generated content as favourably as, or more favourably than, human-written content when seeking social or emotional support. These results indicate that AI-systems may be increasingly accepted as sources of interpersonal guidance. Designers of AI-assisted support systems may benefit from focusing on linguistic features that signal understanding and concern, as these appear central to user perceptions.

# Appendix A

**Control variables**

*Covariates*

The Affinity for Technology Interaction (ATI) scale (Franke et al., 2017) includes 9 items on a six-point Likert scale ranging from 'completely disagree' (1) to 'completely agree' (6), with reverse coding for items three, six, and eight. Items include "I like testing the functions of new technical systems" and "I try to understand how a technical system exactly works".

The AI Attitude Scale (AIAS-4) (Grassini, 2023) spans across four questions using a ten-point Likert Scale from 'not at all' (1) to 'completely agree' (10) without reverse coding. The four items are "I believe that AI will improve my life.", "I believe that AI will improve my work.", "I think I will use AI technology in the future.", "I think AI technology is positive for humanity.".

The Empathy Component Questionnaire (ECQ) (Batchelder et al., 2017) features twenty-seven items ranging from 'strongly disagree' (1) to 'strongly agree' (4) (four-point Likert Scale). Each subcomponent includes two to four inverse-coded items. Items include "I share in other people's emotional experiences" and "I can easily tell when someone is feeling down just by observing their expressions".

# Appendix B

**Task Recall**

Participant were asked to recall the task ones within the first experiment and twice within the second experiment. Specifically, the subjects were provided the following answer options to select from:

I. "Rate empathic advice written by Humans"
II. "Rate empathic advice written by AI"
III. "Rate empathic advice without knowing the author"
IV. "Rate movie script"
V. "Rate descriptions of relationship troubles"

The order of the answer options was randomized.

# Appendix C

Findings for Experiment 2 in table format.

Table S1. *Mean (SD)* human judgments per dependent variable, source and label condition, with variance components (Exp. 2).

| Dependent variable | Source | Label | | ICC | |
|---|---|---|---|---|---|
| | | Human | AI | Rater | Text |
| Overall quality | Human | 3.67 (1.16) | 3.28 (1.26) | 25.5% | 18.1% |
| | AI | 4.13 (0.91) | 3.95 (0.96) | | |
| Cognitive empathy | Human | 3.65 (1.13) | 3.37 (1.16) | 18.1% | 11.4% |
| | AI | 4.22 (0.79) | 4.13 (0.83) | | |
| Emotional empathy | Human | 3.44 (1.23) | 3.12 (1.23) | 31.0% | 10.9% |
| | AI | 3.39 (1.06) | 3.05 (1.16) | | |
| Motivational empathy | Human | 4.04 (0.98) | 3.83 (1.05) | 27.3% | 8.4% |
| | AI | 4.37 (0.74) | 4.19 (0.84) | | |

*Note.* ICC = intra-class correlation (expressed in % of explained variance per random effect)